\documentclass[twocolumn,amsmath,amssymb,floatfix,aps,prx,footinbib,superscriptaddress]{revtex4-1}

\usepackage{amsmath,amssymb,natbib,bm,graphicx,url,epsf,color}
\usepackage[english]{babel}
\usepackage{wasysym}
\usepackage{MnSymbol}
\usepackage{amsmath}
\usepackage{mathtools}
\usepackage{graphicx}
\usepackage{ulem}
\usepackage[colorinlistoftodos]{todonotes}
\usepackage[colorlinks=true, allcolors=blue]{hyperref}
\usepackage{subcaption} 
\usepackage[export]{adjustbox} 
\usepackage{cleveref}

\usepackage{graphicx}
\usepackage[export]{adjustbox}

\begin{document}

\title{Quantum Hall valley splitters and tunable Mach-Zehnder interferometer in graphene}

\author{M. Jo$^*$}
\affiliation{SPEC, CEA, CNRS, Universit\'{e} Paris-Saclay, CEA Saclay, sace91191 Gif sur Yvette  Cedex France}

\author{P. Brasseur$^*$}
\affiliation{SPEC, CEA, CNRS, Universit\'{e} Paris-Saclay, CEA Saclay, 91191 Gif sur Yvette  Cedex France}

\author{A. Assouline}
\affiliation{SPEC, CEA, CNRS, Universit\'{e} Paris-Saclay, CEA Saclay, 91191 Gif sur Yvette  Cedex France}

\author{G. Fleury}
\affiliation{SPEC, CEA, CNRS, Universit\'{e} Paris-Saclay, CEA Saclay, 91191 Gif sur Yvette  Cedex France}

\author{H.-S. Sim$^{\dagger\dagger}$}
\affiliation{Department of Physics, Korea Advanced Institute of Science and Technology, Daejeon 34141, Korea}

\author{K. Watanabe}
\affiliation{National Institute for Materials Science, 1-1 Namiki, Tsukuba, 305-0044, Japan}

\author{T. Taniguchi}
\affiliation{National Institute for Materials Science, 1-1 Namiki, Tsukuba, 305-0044, Japan}

\author{W. Dumnernpanich}
\affiliation{SPEC, CEA, CNRS, Universit\'{e} Paris-Saclay, CEA Saclay, 91191 Gif sur Yvette  Cedex France}

\author{P. Roche}
\affiliation{SPEC, CEA, CNRS, Universit\'{e} Paris-Saclay, CEA Saclay, 91191 Gif sur Yvette  Cedex France}

\author{D.C. Glattli}
\affiliation{SPEC, CEA, CNRS, Universit\'{e} Paris-Saclay, CEA Saclay, 91191 Gif sur Yvette  Cedex France}

\author{N. Kumada}
\affiliation{NTT Basic Research Laboratories, NTT Corporation, 3-1 Morinosato-Wakamiya, Atsugi 243-0198, Japan}

\author{F.D. Parmentier}
\affiliation{SPEC, CEA, CNRS, Universit\'{e} Paris-Saclay, CEA Saclay, 91191 Gif sur Yvette  Cedex France}

\author{P. Roulleau$^{\dagger}$}
\affiliation{SPEC, CEA, CNRS, Universit\'{e} Paris-Saclay, CEA Saclay, 91191 Gif sur Yvette  Cedex France}

\date{\today}


\hspace{2cm}

\begin{abstract}
Graphene is a very promising test-bed for the field of electron quantum optics. However, a fully tunable and coherent electronic beam splitter is still missing. We report the demonstration of electronic beam splitters in graphene that couple quantum Hall edge channels having opposite valley polarizations. The electronic transmission of our beam splitters can be tuned from zero to near unity. By independently setting the beam splitters at the two corners of a graphene PN junction to intermediate transmissions, we realize a fully tunable electronic Mach-Zehnder interferometer. This tunability allows us to unambiguously identify the quantum interferences due to the Mach-Zehnder interferometer, and to study their dependence with the beam-splitter transmission and the interferometer bias voltage. The comparison with conventional semiconductor interferometers points towards universal processes driving the quantum decoherence in those two different 2D systems, with graphene being much more robust to their effect.
\end{abstract}

\maketitle

The simple analogy between the propagation of electrons in a quantum circuit and that of photons in an optics setup has given rise to a field of condensed matter physics called electron quantum optics \cite{Bauerle}, which has led to significant advances in our understanding of the properties of a quantum electrical current. If graphene is considered a very promising material for implementing electron quantum optics experiments\cite{cabart2018}, it is still in its early stage. Indeed, the absence of an intrinsic bandgap has made difficult the development of quantum point contacts and other gate-defined structures similar to those commonly used in GaAs/GaAlAs systems, and central to the implementation of electron quantum optics experiments. The development of tunable electronic beam splitter remains an active research direction, central to the field. The latest realizations of beam splitters in graphene have relied on inducing a bandgap in graphene, either caused by a displacement electric field in bilayer graphene \cite{Danneau2018,Ensslin2020}, or corresponding to the insulating $\nu=0$ quantum Hall state arising at the charge neutrality point (CNP) \cite{ronen2020aharonov,deprez2020tunable}.

In this article, we present electronic beam splitters that directly rely on the valley degree of freedom of graphene.
The principle of our valley beam splitters is based on the theoretical work of \cite{Tworzydlo,Trifunovic}, where the crystalline structure on the corner of a graphene PN junction allows scattering of electrons between PN interface channels, having opposite valley polarization, and quantum Hall edge channels (ECs). We control this scattering by tuning the EC mixing point along the edge of the graphene flake using electrostatic side gates, and we show that the resulting electronic transmission of the valley beam splitters can be reliably changed from zero to near unity, displaying stable and reproducible oscillations with side gate voltage and magnetic field. We develop a theoretical model explaining that these oscillations correspond to a spatial shift of the mixing point over one hexagonal lattice cell, and that their periodicity and amplitude are affected by edge roughness. We then use two valley beam splitters to form, in a controlled fashion, an electronic Mach-Zehnder (MZ) interferometer along the PN junction. Our ability to tune both valley beam splitters up to total reflection allows us to clearly separate the conductance oscillations due to the Aharonov-Bohm flux enclosed by the MZ from the ones due to the valley beam splitters, shedding light on the various oscillatory regimes observed in previous experiments \cite{Morikawa,Wei,Handschin,Makk}. From there we can probe the dependence of the visibility of the MZ interferences with the valley beam splitters transmission and the bias voltage, and compare it to the behavior commonly observed in GaAs/AlGaAs MZ interferometers.

\begin{figure}[h!]
\centering
\includegraphics[trim={6cm 1cm 5cm 1cm},width=0.6\columnwidth,keepaspectratio]{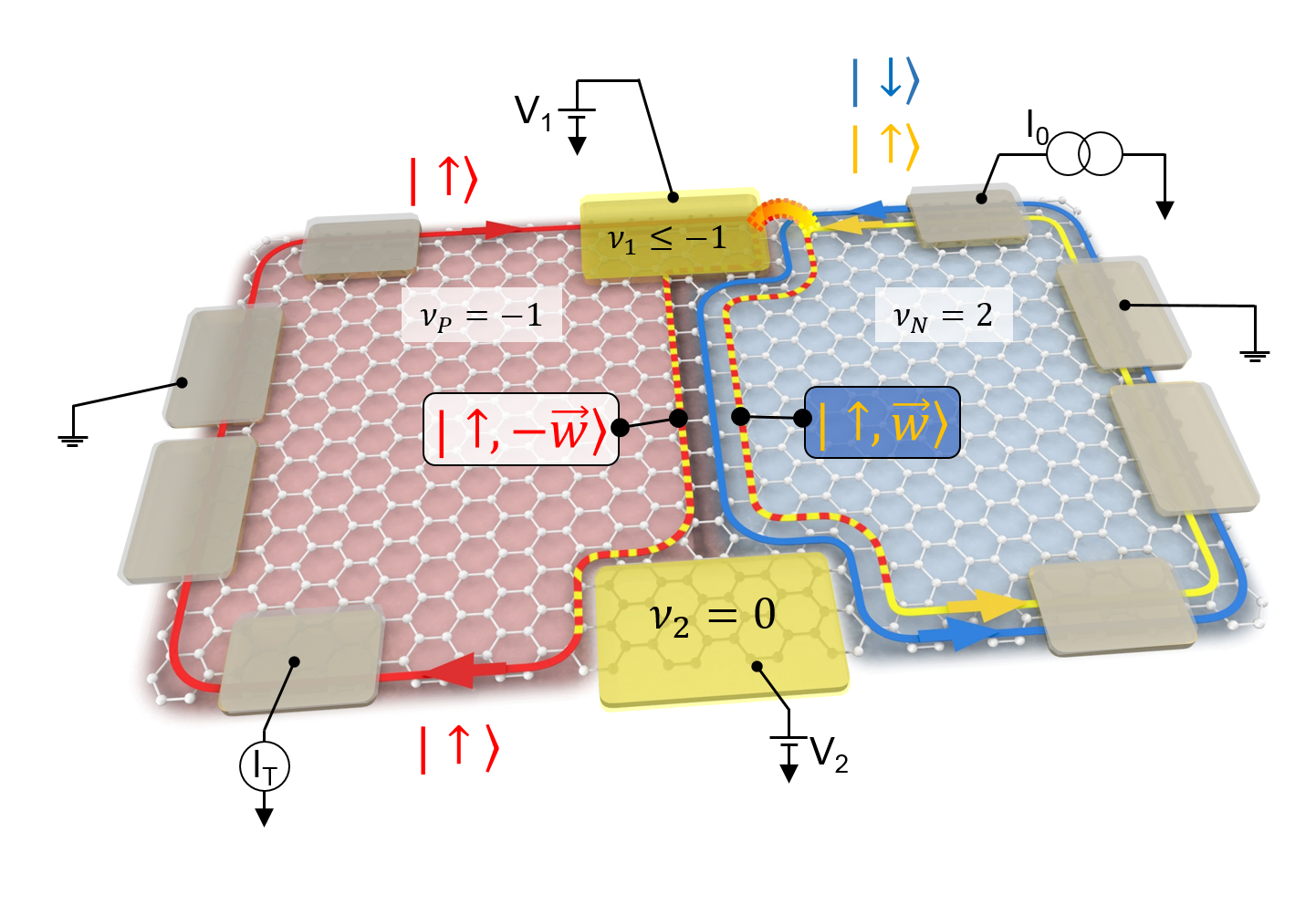} 
\caption{
\textbf{Quantum Hall valley Splitter.} Schematic representation of the PN junction. The N region is depicted in blue, the P one in pink.  Electrons are injected from the upper right ohmic contact (defining an injected current $I_\text{0}$) and transmitted current $I_\text{T}$ is measured at the lower left contact.
}
\label{setup}
\end{figure}


The sample is in a bipolar quantum Hall state, as shown in Fig.~\ref{setup}\cite{Supple1}. 
In the N region the Landau-level filling factor is $\nu_\text{N}=2$ and two ECs of opposite spin ($\uparrow$, $\downarrow$) circulate counterclockwise along the boundary of the sample,
while in the P region $\nu_\text{P}=-1$ and one spin-up channel circulates clockwise.
Along the top edge, the injected current $I_0$ is carried by the two edge channels of the $N$ region.
Half of the current, resulting from spin down carriers, cannot flow to the P region, because of large energy cost for spin flip.
The other half $I_0/2$ with spin up carriers, on which we focus hereafter, can contribute to the transmitted current $I_\text{T}$.

\begin{figure}[h!]
\includegraphics[width=1\columnwidth]{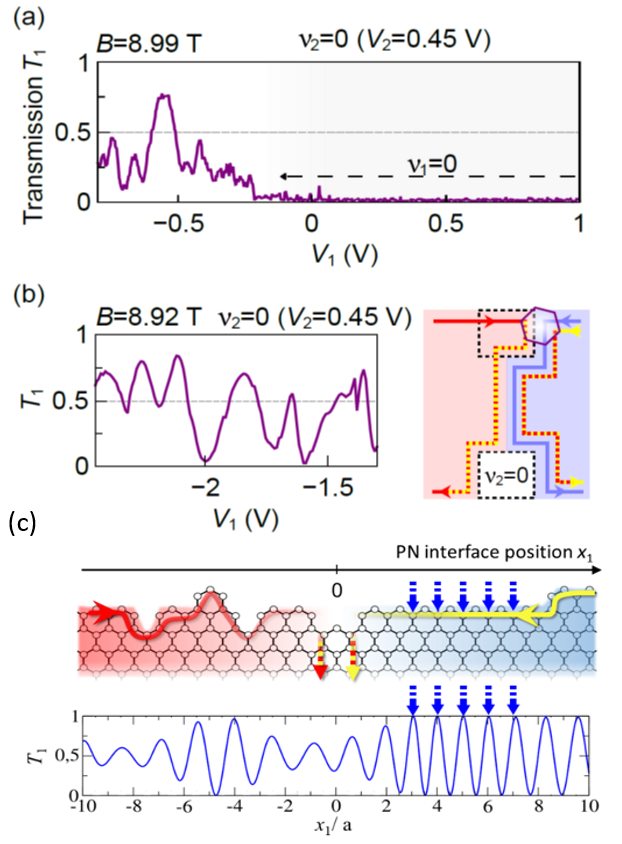}
\caption{
\textbf{Oscillations of the quantum Hall valley splitter.} (a) Measured transmission as a function of the top side gate voltage $V_1$, for $\nu_2=0$ below the bottom side gate. For $\nu_1$=0 below the top side gate (dashed arrow), the injected current is fully reflected. For $\nu_1 \textless 0$, current can be transmitted. (b) Measured transmission $T_1$ of the top beam splitter as a function of $V_1$ for $\nu_1 \le -1$ and $\nu_2=0$. Right panel: sketch of the corresponding edge states configuration. (c) $T_1$ calculated with KWANT as a function of the position $x_1$ of the intersection between the PN interface and the top sample edge. Top panel: edge configuration used in the calculations. Bottom panel: resulting calculated transmission. }
\label{osc}
\end{figure}

The flow of the spin-up current involves splitting into PN-interface channels that have opposite valley isospin \cite{Wei,Wei2018}.
After passing the intersection between the top edge and the PN interface, the current flows along
the P side of the interface with transmission probability $T_1 =  |t_1|^2$
or the N side with reflection probability $|r_1|^2 = 1 - T_1$.
The N-side and P-side currents have opposite valley isospins $\pm \vec{w}$ since valley degeneracy is lifted\cite{Young} under the strong perpendicular magnetic field in the bulk, hence the probability $T_1$ shows the degree of the valley-channel splitting.
This is described by a quantum-mechanical superposition 
\begin{equation}
|\Psi_\text{initial} \rangle =r_1|\uparrow,\vec{w}\rangle+t_1|\uparrow,-\vec{w}\rangle 
\label{initialstate}
\end{equation}
of the N-side spin-down interface state $|\uparrow,\vec{w}\rangle$ and P-side state $|\uparrow, -\vec{w}\rangle$.
Valley-isospin change from that 
of the top edge channel to $\pm \vec{w}$, which is a large momentum change, and occurs due to the atomic structure at the intersection\cite{Trifunovic}. 
We develop a valley splitter in which the transmission probability $T_1$ and the valley isospin are controlled by electrical means.
For this aim, we fabricated two small side gates on the intersections between the graphene physical edge and the PN interface (see Fig.~\ref{setup}). Voltages applied on the side gates modify the profile of the electrostatic potential at both ends of the PN interface.
When the filling factor below a side gate is set to $\nu \leq -1$, the PN junction intersects the physical edge, where the sharp potential change in atomic distance scale facilitates the valley-channel splitting. On the other hand, when the filling factor is set to $\nu=0$, the PN junction intersects an electrically defined edge, where the potential change is smooth and the valley isospin does not change.

\begin{figure*}[t]
    \centering
    \includegraphics[trim={3cm 0cm 0cm 0cm},width=2.0\columnwidth]{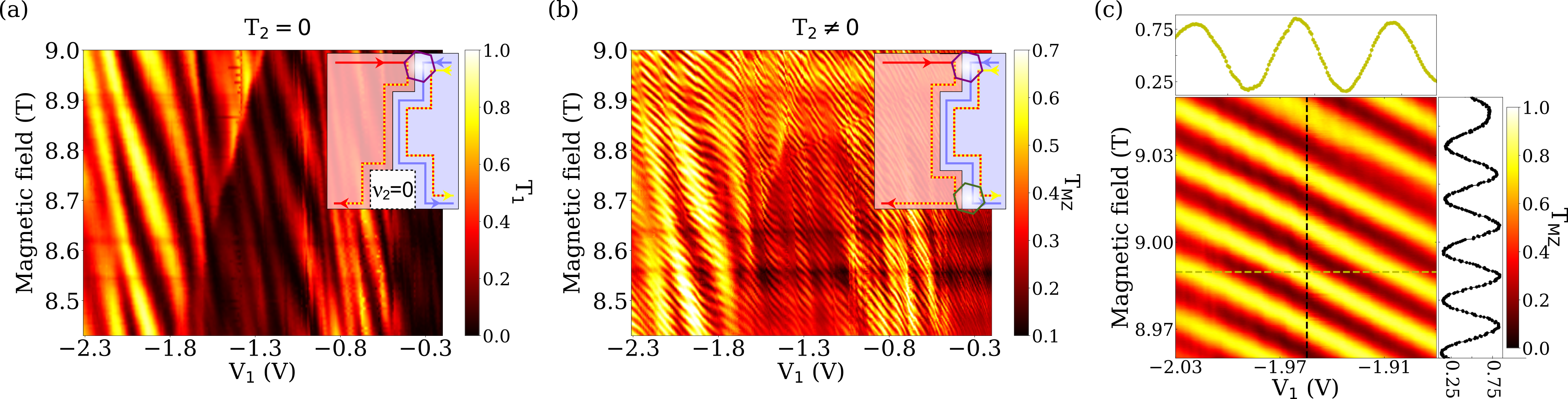}
    \caption{
    \textbf{Oscillations of the Mach-Zehnder interferometer.} 
    (a) Measured $T_1$ as a function of $V_1$ and the magnetic field $B$, with the filling factor below the bottom side gate set to $\nu_2=0$ ($V_2=0.45$~V, see sample configuration in inset). (b) Measured device transmission $T_{MZ}$ in the same range of $V_1$ and $B$, with $\nu_2 \leq -1$, forming a Mach-Zehnder interferometer (inset: sample configuration). (c)  $T_\text{MZ}$ as a function of $B$ and $V_1$ with $T_1 =T_2 \sim 1/2$.}
    \label{osc_Mach}
\end{figure*}


We first demonstrate that the transmission probability $T_1$, defined as $T_1=I_\text{T} / (I_\text{0}/2)$, can be tuned 
by changing the voltage $V_1$ applied on the top side gate (see Figs.~\ref{osc}a and \ref{osc}b).
By setting  $\nu_1 \leq -1$ below the top side gate and $\nu_2 =0$ below the bottom side gate, we expect that the valley-channel splitting occurs at the top intersection but not at the bottom (see the channel splitting shown in Fig.~\ref{setup}). For $V_1>0$ (Fig.~\ref{osc}a), $\nu_1$ is equal to zero, and the PN junction only intersects electrostatically defined edges. This fully suppresses valley-channel splitting, and the measured transmission vanishes. Conversely, for $V_1<0$, $\nu_1\leq -1$ and the PN junction intersects the top physical edge: valley-channel splitting now occurs at the top intersection only, leading to a finite transmission. We show in Fig.~\ref{osc}b that the transmission probability $T_1$ can be tuned between zero and almost unity by changing the voltage $V_1$ (importantly, $T_1$ can also be tuned by changing the magnetic field, see \textit{e.g.} Fig.~\ref{osc_Mach}a).
The period of the $V_1$ dependence is $\Delta V_1 \sim 100$ mV on average.
Combining this and the Aharonov-Bohm oscillation in Fig.~\ref{osc_Mach}b, we discuss later that $\Delta V_1$ makes the PN-interface shift by $\sim 1$ nm below the top side gate.
The period of the $B$ dependence is $\Delta B_1 \sim 300$ mT, which corresponds to $\sim 0.2$ nm change of the magnetic length around $B = 9.2$ T.
These lengths are comparable with the interatomic distance 0.142 nm of the pristine graphene and the period of atomic edge structures (e.g., 0.246 nm of the zigzag edge), and much shorter than the scale of spatial variation of gate-voltage induced electrostatic potential.
This strongly suggests that the transmission $T_1$ is controlled by the atomic structure at the top intersection.
The PN-interface shift is estimated from the experimental data, without the help of dedicated numerical simulations.
This direct estimation is possible with independent control of the top and bottom side gates and could not be done in previous works\cite{Wei,Handschin}. 
We also measure the transmission probability $T_2 = I_\text{T} / (I_0/2)$ of the junction as a function of the voltage $V_2$ applied on the bottom side gate, when the filling factor below the top (bottom) side gate is $\nu_1 = 0$ ($\nu_2 \leq -1$). In this case, valley-channel splitting occurs not at the top intersection but at the bottom; the spin-up injected current flows along the N side of the PN interface after passing the top intersection, and then flows, after passing the bottom intersection, to the bottom left edge with probability $T_2$ or to the bottom right with probability $1-T_2$. Reproducible irregular oscillations in $T_2$ are also observed (see Fig. S10).


\begin{figure*}[t]
    \centering
    \includegraphics[width=2.\columnwidth]{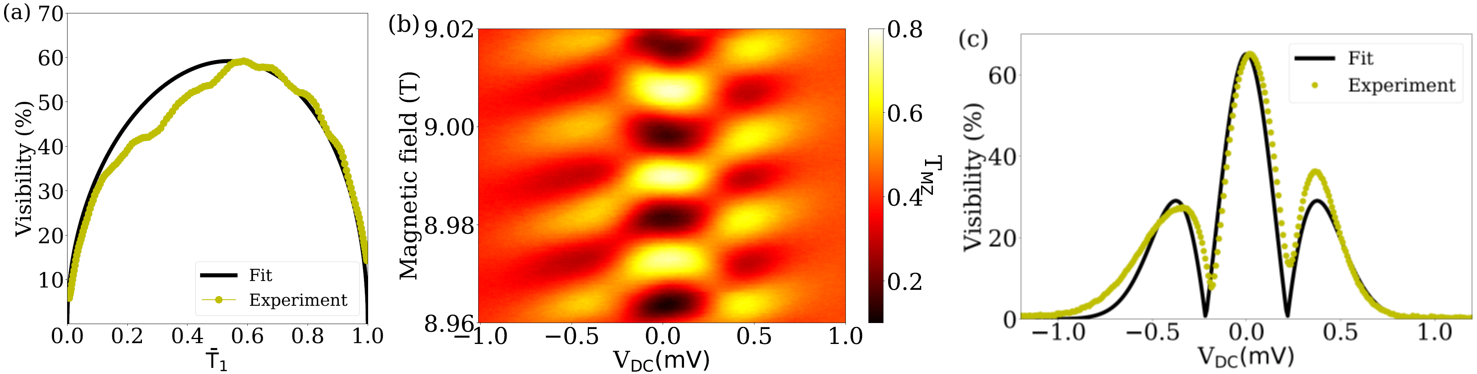}
    \caption{
    \textbf{Transmission and Energy dependence of the Mach-Zehnder visibility.} (a) Interference visibility as a function of the normalized $\bar{T}_1$ (see supplemental materials).
    The visibility (yellow dots) of the experimentally measured oscillation of $T_\textrm{MZ}$ and the curve of $\alpha\sqrt{\bar{T_1}(1-\bar{T_1})}$ where $\alpha$ is an adjustable parameter show good agreement. (b) $T_\text{MZ}$ as a function of the DC bias $V_\text{DC}$ (applied to the upper right ohmic contact for the injection of $I_0$) and the magnetic field.  
    (c) Measured visibility (yellow dots) as a function of $V_\text{DC}$. 
    Computed visibility (black solid line) based on gaussian phase fluctuations is in agreement with the experimental data.}
    \label{T_dep}
\end{figure*}

The irregular but reproducible oscillations in the transmissions are attributed to the atomic structure of the intersection between the PN interface and the sample's edge. This is supported by our numerical calculations (see Fig.~\ref{osc}d, and Appendix E of \cite{Supple}) using KWANT\cite{Groth}. The transmission $T_1$ is computed in the vicinity of the top edge while varying the position $x_1$ of the top intersection. This corresponds to a change in the top side gate voltage $V_1$. For a clean zigzag edge ($x_1>0$ in Fig.~\ref{osc}d), $T_1$ shows regular oscillations as a function of $x_1$, with a period $a=0.246$~nm matching the atomic-structure period of the zigzag edge (blue arrows in Fig.~\ref{osc}d). 
The oscillation is irregular when roughness is added along the top edge ($x_1<0$), but it has periods not much modified from the clean case. Disorder yields similar results, discussed in Appendix E. Our simulations are qualitatively compatible with the experimental data.

By tuning the filling factor below both the top and bottom side gates to $\nu \leq  -1$, we have two valley splitters in series. Combining them, we further characterize and control the valley splitting and valley isospin. This is illustrated in Fig.~\ref{osc_Mach}, where we compare the device transmission $T_1$ when the filling factor below the bottom side gate is set to $\nu_2=0$ (Fig.~\ref{osc_Mach}a) to the transmission $T_\text{MZ} = I_\text{T} / (I_0/2)$ when $\nu_2\leq -1$  (Fig.~\ref{osc_Mach}b). Strinkingly, we observe regular, smaller scale oscillations in $T_\text{MZ}$ ($\Delta B^{\text{MZ}} \sim 25$ mT, and $\Delta V_1^{\text{MZ}} = 50$ mV) that are superimposed to the larger scale oscillations of $T_1$ shown in Fig.~\ref{osc_Mach}a.
These oscillations are naturally interpreted as interference fringes of a Mach-Zehnder interferometer where the interface channels $|\uparrow, \pm \vec{w} \rangle$ constitute the two interferometer arms and the valley splitters behave as the beam splitters.
From the Aharonov-Bohm phase $\phi_\text{AB} = 2 \pi BA / \Phi_0$, $\Phi_0=h/e$ being the flux quantum, we get the interferometer area $A = \Phi_0 /\Delta B^{\text{MZ}} = 0.15 \, \mu \text{m}^2$ and the spatial separation of 110 nm between the two interface channels, given the length 1.5 $\mu$m of the PN interface. The separation results from electron-electron interactions\,\cite{Wei}.

The period $\Delta V_1^{\text{MZ}} = 50$ mV indicates that the interferometer area $A$ is tuned by $V_1$. The gate voltage $V_1$ generates potential to the P side and the N side asymmetrically, resulting in change $\delta s$ of the spatial separation between the interface channels and shift $\delta d \sim \delta s / 2$ of the PN interface below the top side gate  (as depicted in Fig.~\ref{osc_Mach}d).
We estimate $\delta s \sim 1$ nm, hence $\delta d \sim 0.5$ nm, for $\Delta V_1^{\text{MZ}} = 50$ mV from
the Aharonov-Bohm flux $\delta s L_s B = \Phi_0$ of the area $\delta s L_s$ at $B = 9.2$ T with the side-gate length $L_s = 450$ nm. This implies that $\Delta V_1 = 100$ mV applied on the top side gate voltage $V_1$ enables to shift the PN interface by a distance of  $\delta d$ $\sim 1$ nm, supporting that the dependence of $T_1$ on $V_1$ in Fig.~\ref{osc} is related with the valley isospin.
The two measurements in Figs.~\ref{osc_Mach}a and b are perfectly consistent and confirm our interpretation.
We remark that independent control of the two valley splitters is crucial for characterization of the valley splitters and the Mach-Zehnder interferometers, as there are multiple periodicities of different origin such as $\Delta B_1$, $\Delta B^{\text{MZ}}$, $\Delta V_1$, and $\Delta V_1^{\text{MZ}}$. This was not achieved previously\,\cite{Wei,Handschin}.

We provide a theoretical understanding for the transmission probability  $T_\text{MZ}$.
The valley superposition in Eq.~\eqref{initialstate} further evolves, gaining the Aharonov-Bohm phase, into a state $|\Psi_\text{final} \rangle =r_1|\uparrow,\vec{w}\rangle+t_1 e^{i \phi_\text{AB}}|\uparrow,-\vec{w}\rangle$ at the interface bottom.
This happens when there is no atomic defect along the PN interface (as in our experiment, AB oscillations being extremely robust over a wide parameter range of the magnetic field and gate voltage) so that the valley isospin $\pm \vec{w}$ does not change.

Therefore, the valley superposition is further engineered by using the Aharonov-Bohm effect.
On the other hand, the bottom valley splitter is characterized by a scattering eigenstate\,\cite{Trifunovic,Tworzydlo}  $|\widetilde{\Psi} \rangle$ of the interface that perfectly goes to the left-moving bottom edge channel; the state orthogonal to  $|\widetilde{\Psi} \rangle$ moves to the right-moving channel. We write this state as $|\widetilde{\Psi} \rangle =t_2|\uparrow,\vec{w}\rangle+r_2 |\uparrow,-\vec{w}\rangle$ with amplitudes $r_2$ and $t_2$ satisfying $|r_2|^2 + |t_2|^2 = 1$, choosing the interface states $|\uparrow, \pm \vec{w} \rangle$ as basis states.
The transmission probability is obtained as   $T_\text{MZ} = |\langle\widetilde{\Psi}|\Psi_\text{final}\rangle|^2 = |r_1 t_2^*  + t_1 r_2^* e^{i \phi_\text{AB}}|^2$,
\begin{equation}
T_\text{MZ} =  |r_1t_2|^2+|r_2t_1|^2+2|r_1t_1r_2t_2|\text{cos}(\phi_{\text{AB}}+\phi),
\label{interference}
\end{equation}
where $\phi = \text{arg} (t_1 t_2 r_1^* r_2^*)$ can include the dynamical phase.
This has the standard form of the Mach-Zehnder interference, but includes the effect of the valley isospin.

We study the interference visibility $(T_\text{MZ,max}-T_\text{MZ,min})/(T_\text{MZ,max}+T_\text{MZ,min})$, where $T_\text{MZ,max(min)}$ is the maximum (minimum) value of the oscillation of $T_\text{MZ}$.
To obtain the largest visibility, we first set the bottom side gate to $\nu_2 = 0$ and tune $V_1$ to get
$T_1 \sim 1/2$. Next we set the bottom side gate to $\nu_2 \leq -1$ and finely tune $V_2$ to reach the maximum visibility. According to Eq.~\eqref{interference}, we expect the visibility $\propto \sqrt{T_1 (1-T_1)}$. This behavior is found in Fig.~\ref{T_dep}a. The excellent agreement between the experimental visibility and the $\sqrt{T_1 (1-T_1)}$ law obtained with the independent measurement of $T_1$ supports the formation of the Mach-Zehnder valley interferometer and confirms the coherence of the valley superposition in Eq.~\eqref{initialstate}.
The analysis indicates that the valley isospin direction of the superposition $|\Psi_\text{final} \rangle$
is tuned over the range of $|t_1|^2 =$ 0.018 - 0.812 and $\phi_\text{AB} =$ 0 - $2\pi$ with changing $V_1$ and $B$.

 In the last section, we address the coherence properties of the valley-split states propagating along the PN interface when electrons are injected at higher energy. In Fig.~\ref{T_dep}b, $T_{\text{MZ}}$ is represented as a function of the applied bias voltage and the magnetic field. We observe a lobe-type structure, accompanied by a phase shift of $\pi$ when the visibility is canceled. This behavior is a rather common observation in MZ interferometers realized in GaAs/AlGaAs heterostructures\,\cite{Neder, Roulleau_2007}. While its microscopic origin remains unknown, it is widely thought to stem from interaction effects that lead to a Gaussian averaging of the interferometer phase as the bias voltage is increased. The phase can be written $\varphi=\phi_{AB}+\phi$ (see Eq.\ref{interference}), and its variance $\langle\delta\varphi^2\rangle=V_\text{DC}^2/V_\text{lob}^2$ where $\langle\rangle$ denotes the average over the phase distribution. $V_\text{lob}$ reflects how electronic interactions induce decoherence in the interferometer, and is larger for robust quantum interferences. The oscillatory part of the transmitted current is written $I_\text{T}=T_\text{MZ}\times(I_0/2)\times  e^{-V_\text{DC}^2/2V_\text{lob}^2}$. Since we measure the differential conductance $dI_\text{T}/dV_{DC}$, this leads to a visibility $\propto e^{-V_\text{DC}^2/2V_\text{lob}^2}|1-V_\text{DC}^2/V_\text{lob}^2|$ (black solid line in Fig.~\ref{T_dep}c) \cite{Roulleau_2007}. Whereas typical reported values in GaAs/AlGaAs heterostructures give $V_\text{lob}\sim 20$~$\mu$eV \cite{Neder,Roulleau_2007,Bieri}, we observed a much larger value $V_\text{lob}=217$ $\mu$eV \cite{Supple2}. Theoretically, most of the models developed to account for side lobe patterns in GaAs/AlGaAs heterostructures rely on the partitioning induced by the first beam splitter\,\cite{Levkivskyi,Youn}. Experimentally a clear dependence of the lobe structure has been observed in GaAs/AlGaAs heterostructures as a function of the transmission of the first beam splitter\,\cite{Bieri}. Unexpectedly, we do not observe such a dependence in graphene (see Fig.\,S9 in the supplemental material).

To conclude, we have demonstrated that monolayer graphene in the quantum Hall regime is a promising platform to perform electron quantum optics experiments. We have first  characterized a quantum Hall valley splitter showing that the valley mixing between two opposite valley isospin edge states can be finely tuned. Coherence of this valley beam splitter is discussed by implementing it in a PN-junction to define a fully tunable electronic Mach-Zehnder interferometer. This new type of quantum-coherent valleytronics
devices should allow to envision
two-valley-state operations and entanglement schemes\cite{Samuelsson}. The demonstration of quantum coherent valleytronics opens another large field of quantum transport devices in graphene including quantum dots, Kondo impurities, and various interferometers such as Fabry-Perot resonators. \\

\acknowledgments
We warmly thank Thierry Jolicoeur and Mark Goerbig for enlightening discussions, as well as P. Jacques for technical support. This work was funded by the ERC starting grant COHEGRAPH, the CEA, the French Renatech program, ``Investissements d\textsc{\char13}Avenir" LabEx PALM (ANR-10-LABX-0039-PALM) (Project ZerHall), and by the EMPIR project SEQUOIA 17FUN04 co-financed by the participating states and the EU’s Horizon 2020 program. It is also supported by Korea NRF via the SRC Center for Quantum Coherence in Condensed Matter (GrantNo.2016R1A5A1008184).

$^{\dagger}$ main corresponding author
$^{\dagger\dagger}$co-corresponding author
$^{\star}$ equal contribution

\vspace*{50mm}

\end{document}